\documentclass[12pt,preprint]{aastex}

\shorttitle{Supernovae in Arp 220}
\shortauthors{Lonsdale et al.}
\usepackage{lscape}
\usepackage{pdflscape}

\begin{document}

\title{VLBI Images of 49 Radio Supernovae in Arp 220}

\author{Colin J. Lonsdale}
\affil{MIT Haystack Observatory, Westford MA 01886, USA}
\email{cjl@haystack.mit.edu}

\author{Philip J. Diamond and Hannah Thrall}
\affil{Jodrell Bank Observatory, Macclesfield, SK11 9DL, UK}
\email{pdiamond@jb.man.ac.uk, hthrall@jb.man.ac.uk}

\author{Harding E. Smith}
\affil{CASS, U.C. San Diego, 9500 Gilman Dr., La Jolla CA 92093-0424, USA}
\email{hsmith@ucsd.edu}

\and

\author{Carol J. Lonsdale}
\affil{Infrared Processing and Analysis Center, MS 100-22, Pasadena, CA
91125, USA}
\email {cjl@ipac.caltech.edu}

\begin{abstract}
We have used a Very Long Baseline Interferometry (VLBI) array at 18cm
wavelength to image the nucleus of the luminous IR galaxy Arp 220 at
$\sim$1 pc linear resolution, and with very high sensitivity.  The
resulting map has an rms of 5.5 $\mu$Jy/beam, and careful image analysis
results in 49 confirmed point sources ranging in flux density from 1.2 mJy
down to $\sim$60 $\mu$Jy.  Comparison with high sensitivity data from 12
months earlier reveals at least four new sources.  The favored
interpretation of these sources is that they are radio supernovae, and if
all new supernovae are detectable at this sensitivity, a resulting
estimate of the supernova rate in the Arp 220 system is 4$\pm$2 per year.
The implied star formation rate is sufficient to power the entire observed
far-infrared luminosity of the galaxy.  The two nuclei of Arp 220 exhibit
striking similarities in their radio properties, though the western
nucleus is more compact, and appears to be $\sim$3 times more luminous
than the eastern nucleus.  There are also some puzzling differences, and
differential free-free absorption, synchrotron aging and expansion losses
may all be playing a role.  Comparison with the nearby starburst galaxy
M82 supports the hypothesis that the activity in Arp 220 is essentially a
scaled-up version of that in M82.  
\end{abstract}

\keywords{galaxies: active --- infrared: galaxies --- radio
continuum: galaxies}

\section{Introduction}
A starburst is an intense period of star formation within a galaxy that
cannot be sustained over its lifetime, due to the huge amount of gas
needed to fuel the process.  Typical starburst lifetimes are 10$^8$ to
10$^9$ years. Starbursts can occur on two scales: centralized in the
galactic nucleus (a nuclear starburst) or throughout the entire galaxy (a
global starburst). In order to trigger a starburst, there must be a
plentiful supply of gas to the centre of the galaxy, which occurs when a
large-scale disturbance to the angular momentum of the central region
causes the infall of enough material. In a global starburst this
disturbance is usually associated with the collision (and possible merger)
of two galaxies. In a nuclear starburst (such as M82) the disturbance may
be due to the effects of shock fronts caused by a stellar bar, a tidal
interaction with another galaxy, or perhaps an active galactic nucleus
(AGN). The most luminous starbursts appear to all be triggered by
interactions with other galaxies.  Smith, Herter and Haynes (1998) studied
a sample of 20 radio-luminous starburst galaxies. They found that all 20
galaxies were mergers, interacting pairs, or members of groups or
clusters, suggesting that all the starbursts in the sample were most
likely triggered by interactions with other galaxies.

It has become clear that a ``pan-spectral" approach to observations is
necessary to comprehensively study starburst galaxies, combining
observations taken over as large a wavelength range as possible, and
supported by theoretical advances and modeling. However, particularly as
the more compact and luminous nuclear starbursts are enshrouded in dust
clouds which are optically thick at shorter wavelengths, long wavelength
data are uniquely able to peer through the curtain of extinction and
derive the basic properties of the stellar population. In some cases, IR
data is sufficient, but where even IR radiation is strongly absorbed, it
is necessary to turn to the mm and radio domains.

The IR luminosity is a widely used diagnostic of starburst activity. This
strong emission comes from dust grains presumed to be heated by photons
from young stars formed in the starburst. Assuming that most of the energy
in photons emitted by young stars eventually results in heating the dust
(which then re-radiates it as black-body thermal continuum in photons),
then star formation rates for stars more massive than 5M$_\odot$ can be
estimated from the far-IR luminosity (Cram et al 1998). Such stars
dominate the total stellar luminosity, and end their lives as supernovae.
Supernova rates in starbursts can be directly related to formation rates
of massive stars, and by assuming an initial mass function (IMF), the
luminosities of the starbursts.  

Studies of radio supernovae (RSNe) in starbursts are unlike studies
in our own Galaxy. The sample of RSNe in a starburst is well-defined
and largely free from selection effects, and all the sources can be
considered to be at the same distance. Therefore, nearby starbursts
can be considered as laboratories in which to investigate, both
statistically and individually, the properties of a well-defined
sample of RSNe and supernova remnants (SNR).

Arp 220, at a distance of 77Mpc and with an infrared luminosity of
L$_{IR} \sim 1.3 \times 10^{12}$L$_\odot$ (Soifer et al., 1987) is one of
the closest and best studied Ultra Luminous InfraRed Galaxies, or ULIRGs.
It is, like most ULIRGs a merger of two gas-rich galaxies; Arp 220 has a
pair of radio nuclei separated in roughly the E-W direction by
approximately 370 pc (Norris, 1988, Rovilos et al, 2003, Rodriguez-Rico et
al., 2005).  The two nuclei are also clearly seen by high-resolution Keck
imaging in the mid-infrared (Soifer et al., 1999, 2002).  The western
nucleus is brighter than the eastern one, though the flux ratio is
frequency dependent.

Smith et al (1998a) made 18cm continuum observations of Arp 220, in which
they found a number of unresolved sources in the western nucleus which
they interpreted as luminous radio supernovae. They suggested that this
interpretation is consistent with a simple starburst model for the IR
luminosity of Arp 220 that has a star formation rate of 50-100
M$_\odot$/yr and a luminous supernova rate of 1.75-3.5/yr. The supernovae
observed are the most clearly detected RSNe in a distant galaxy. Smith et
al suggested, based on the extreme flux densities, that the objects might
be Type IIn supernovae.  This type of supernova is optically identified by
a narrow component in its spectrum and is thought to be interacting with a
dense circumstellar medium, a result of the high mass loss rates of the
progenitor stars.  Rovilos et al (2005) obtained multi-epoch high
resolution data of Arp 220, and concluded that the light curves had been
surprisingly stable over a period of more than five years. The long decay
time of the 18cm light curves of these sources does not agree with the
simple model of Type IIn supernovae (Smith et al. 1998a). The observations
discussed in this paper are part of a subsequent high-sensitivity
monitoring programme of Arp220.

\section{Observations and Data Reduction}
We observed Arp220 on 9 November 2003 with a global array of the largest
telescopes on Earth. The observations were part of a long-term monitoring
campaign aimed at studying the changes in the compact structure of both
the continuum and OH maser emission in Arp220. 18 antennas took part in
the experiment, including the 10 VLBA antennas, the phased VLA, the GBT,
Arecibo, and 5 antennas of the European VLBI Network (EVN). All antennas
performed nominally except for the phased-VLA; it was unable to maintain
phase coherence due to the presence of interference, and so all phased-VLA
data were flagged.

The observations covered a period of almost 21 hours. Data were taken in 2
polarizations and in four 8 MHz wide intermediate frequency (IF) bands
covering the frequency range 1633.99 -- 1665.99 MHz and with an aggregate
data rate of 256 Mbit per second. The strong redshifted OH maser emission
from Arp220 was centered in the lowest of the four IF bands. The line-free
portions of the lowest IF, plus the upper 24 MHz from the other 3 IFs were
dedicated to observing the continuum emission from Arp220. The data were
recorded on tape and shipped to the VLBA correlator in Socorro, New
Mexico. There, two correlator passes were done, one with 512 channels
across the 8 MHz of the bottom IF containing the OH line, and a second
with 32 channels per IF across all four IFs.  The correlated data were
sent to Jodrell Bank Observatory for subsequent data processing.

Due to the presence of strong and compact OH maser emission (Lonsdale et
al. 1994, 1998), which could be used for phase calibration, there was no
need for phase referencing. Therefore, the observing scheme was
straightforward, with most time being spent on Arp220 and with occasional
15 minutes scans on calibrator sources for the purposes of tracking the
residual delays and determining the bandpass corrections. The calibrator
sources used for this purpose were 1613+341 and 1516+193. The J2000
coordinates assumed for Arp220 were: RA = 15h 34m 57.225s, Dec =
23$^{\circ}$ 30' 11.564".

All data calibration was carried out using the 15DEC01 release of
NRAO's AIPS software and used standard procedures (Diamond, 1995), to
summarize: the data were corrected for digital effects and those of
parallactic angle rotation; the gross residual delay errors in each
observing band and the phase offsets between the bands were
determined; amplitude calibration was performed using the system
temperature and gain curves supplied by the observatories; any time
variable residual delay offsets and the full complex bandpass solution
for each antenna were determined; the residual antenna-based phase
delay rates of the strongest channel in the OH spectrum was measured at 5
minute time intervals. All of these calibration factors were then
applied to the data to produce a data set coherent in time and
frequency. Then, an iterative loop of self-calibration and imaging was
performed on the OH reference channel to determine and correct for the
short timescale (~1 minute) antenna-based phase errors; the
corrections thus derived were applied to all the frequency bands
thereby ensuring that the positional registration of the continuum
emission, when imaged, was anchored to the position of the peak of the
strong line emission.

The continuum emission in Arp220 is problematic from a VLBI imaging
perspective in that the structure has a very different character depending
upon the spatial resolution with which it is viewed. As we have previously
shown with MERLIN (Rovilos et al 2003), on arc-second scales we see all of
the expected $\sim$280mJy flux from Arp 220 (Baan, Wood \& Haschick 1982).
About 200 mJy of this is concentrated in two main components separated by
1 arcsecond almost east-west, which are surrounded by more extended
emission.  However, our early VLBI observations (Smith et al., 1998a)
showed that on baselines beyond a few hundred kilometres most of the
MERLIN-scale structure was resolved and we only see a few percent of the
flux concentrated in a number of point sources.  Therefore, in producing
the continuum image from this dataset, which suffers from poor (u,v)
coverage on short baselines, we must take care when imaging. We found that
the best approach was to simply eliminate the short baselines and the
diffuse emission detected on them, by ignoring all data within a (u,v)
range of 5000k$\lambda$, i.e. projected baseline lengths of less than
900km.  This results in only minimal loss of sensitivity to compact
features, because the largest apertures in the telescope array tend to be
at the periphery of the array, and tend not to participate in many short
baselines.

Regions of the final image centered on the two nuclei of Arp 220 are shown
in Figure 1.  The rms noise across the image varies somewhat with location
due to dynamic range limitations in areas with a high density of compact
sources.  However, we find that in the emptiest areas of the image the rms
noise is 5.5 $\mu$Jy/beam, and close to the NW region of compact structure
it approaches 9 $\mu$Jy/beam: this image is, therefore, the most sensitive
VLBI image obtained to date by some margin.  Many point sources are
clearly visible.  Due to the variable noise levels and the large number of
faint source candidates, objective criteria for identifying and measuring
sources were required, as described in the next section.

\section{Image Analysis and Results}

The focus of this investigation is the population of point sources in the
two nuclei of Arp 220.  For a reliable and complete source list, we need
objective and robust criteria for source detection.  A careful treatment
is required, because the deconvolution step of image generation, via the
CLEAN algorithm, is highly non-linear in nature, contributing to strongly
non-Gaussian noise statistics in the final image; another factor that may
influence the statistics is residual poorly constrained diffuse continuum
emission.  The statistic we need to develop for each candidate source is a
{\em probability of false detection}, $\Psi$, determined empirically from
analysis of noise statistics as a function of position within the image.

As noted above, the rms fluctuation level varies across the image.  Rms
values have been measured in the immediate vicinity of each source
candidate C, which we refer to as {\em local variances} and denote by
$\sigma_C$.  For a measured flux density $S_C$ of a candidate source, we
define an effective signal to noise ratio $\eta$, such that $S_C =
\eta.\sigma_C$.  We can then estimate the probability of false detection
$\Psi_C$ for source candidate C by reference to an empirically determined
$\Psi(\eta)$.  Construction of our source list is then accomplished by
measuring all candidates down to some low flux density level, and then
applying a cutoff threshold in $\Psi_C$ (which is equivalent to a
threshold in $\eta$).

In order to determine $\Psi(\eta)$ in a manner that is representative of
the image areas in which candidate sources are being measured, we removed
all source candidates from the image, along with the few areas immediately
adjacent to brighter sources that suffer from obvious sidelobe artifacts.
We then selected two areas, one enclosing the eastern nucleus and one the
west nucleus.  We also analyzed a comparison region between the two nuclei.
Histograms of the candidate-free pixel values were constructed, and
probability distributions of local rms levels were derived (Figure 2).  It
can be seen that despite the factor of ~1.4 difference in rms levels
between east and west, the {\em shape} of the probability distributions is
very similar in the two nuclei, and quite similar in the comparison
region.  All three distributions are also strongly non-Gaussian, as
illustrated in the diagram.  While we show only positive pixel values
here, the distributions are closely symmetric around zero, which implies
that the departure from Gaussian statistics is not due to a population of
weak sources below the detection limit, and is instead a product of the
image reconstruction algorithms.

We have extrapolated the probability function for the regions enclosing
the two nuclei via a least-squares linear fit to the data from both
regions (dashed black line).  We have then used this fit to estimate
$\Psi(\eta)$.

Note that pixels on the image are not independent on the angular scale of
the array resolution.  In addition, the observed strong clustering of
candidate sources around the two nuclei defines a limited region of sky
over which our search for sources is to be conducted.  Therefore, while
the full image contains $\sim$2 million pixels, the search is conducted
over only $\sim 2 \times 10^4$ independent resolution elements.  It is the
probability of false detection within this number of independent image
samples to which $\Psi(\eta)$ must refer.

For a given candidate source with SNR $\eta_C$ (i.e. flux density divided
by local rms), the probability that any given pixel value $\eta_P$ is less
than this is given by $P(\eta_P < \eta_C) = 1 - P(\eta_P > \eta_C)$, where 
log$[P(\eta_P > \eta_C)]$ is the quantity on the vertical axis of Figure
2.  By using a least-squares linear fit to the data from the two nuclei
displayed in Figure 2, we model this function as log$[P(\eta_P > \eta_C)]
= -0.446 \eta_C - 0.788$, denoted by the dashed black line.  The
probability that all pixels searched are below $\eta_C$ is $[1 - P(\eta_P
> \eta_C)]^n$, where $n$ is the number if independent resolution elements
searched ($2 \times 10^4$ in this case), and we have

\[ {\rm log} ( 1 - \Psi_c) = n \, {\rm log} [1-P(\eta_P > \eta_C)]  \]

and the total expected number of false detections is then given by

\[ N_{false} \sim \sum_C \Psi_C \]

In order to achieve $N_{false} < 1$, we find that we must omit candidate
sources with $\eta_C < 9.5$.  The result is a list of 47 sources, with the
expectation of less than one false positive.  There are also 4 candidate
sources with $9 < \eta_C < 9.5$, of which 2 have faint but unmistakable
counterparts on an image from 12 months earlier (not yet published), and
in which we therefore have high confidence.  Our final source list thus
contains 49 objects, listed in Table 1, which are the basis for the
ensuing discussion.

We have adopted a naming convention of E{\em n} and W{\em n} for sources
in the east and west nuclei respectively, where {\em n} is a sequential
candidate number, arbitrarily in order of decreasing Right Ascension.
Not all candidates met our detection criteria, so the source numbers in
Table 1 are not sequential.  In future papers, newly discovered sources
will be assigned higher sequential numbers.

As a footnote from this analysis of image statistics, it is an interesting
fact that in all three regions analyzed (east nucleus, west nucleus and
control region), the data values below a SNR of $\sim$1 closely follow a
Gaussian distribution, with equivalent rms values just below 5
$\mu$Jy/beam in all three cases.  The location-dependent departures from
Gaussian distributions represent a high kurtosis parameter, and a
`leptokurtic' distribution.  The consistency of the low SNR Gaussian fits
strongly suggests that the underlying thermal noise level on the map is
$\sim 4.8 \mu$Jy/beam, and that the rest is sidelobe noise propagating
through the complex, non-linear imaging process.  Our results may provide
a cautionary note to others seeking to interpret low-level features in
similar images.

\section{Discussion}

\subsection{General Observations}

Figure 1 shows the overall structure of the continuum emission from Arp220
as observed with VLBI.  The sensitivity level achieved in this image is
significantly better than earlier efforts (Smith et al., 1998a; Rovilos et
al., 2003), and represents the current state of the art in high sensitivity
radio imaging.  As a result of the sharply reduced noise levels, we have
been able to detect 49 continuum point sources in Arp 220 (29 in the west
nucleus, and 20 in the east).  From the current observations, we note the
following pertinent features of the structure:

\begin{itemize} 
\item{} All 49 sources appear unresolved to our array, and
typical upper limits on their linear extent are less than 1 parsec.
\item{} The sources are clustered, with 22 of the 29 western sources
falling within a 0.25$\times$0.15 arcsecond region, and 14 of the 20
eastern sources falling within a 0.3$\times$0.2 arcsecond region.  In both
nuclei, sources falling outside these compact regions tend to be weaker.
\item{} The total flux density accounted for in the point sources is 2.0
mJy in the east and 9.8 mJy in the west.  The total 18cm continuum flux
densities of the two nuclei, for comparison, are 92 and 111 mJy
respectively, with another $\sim$80mJy on larger scales (Rovilos et al.,
2003). {\em The VLBI-scale emission accounts for only a few percent of the
total.}
\item{} The diffuse emission, which dominates the total 18cm emission from
Arp 220, is generally resolved out by our array (but see discussion in
section 2).  Based on the clustering, we estimate that the Rovilos et al.
(2003) MERLIN peak flux density may have up to a 15\% contribution from
the compact sources in the west, and less than 2\% in the east.  
\item{} Sources in the east nucleus are systematically weaker than those
in the west.  Based on the expanded size of the sample now available, this
is a robust result.  {\em The apparent luminosity function of the sources
is markedly different between the two nuclei.} 
\item{} The source distributions in the two nuclei mirror the properties
of the diffuse emission, namely stronger and more compact in the west,
weaker and less compact in the east.  Furthermore, the centroids of the
source clusters in west and east closely match the relative locations of
the two nuclei as seen at 6cm by MERLIN (see Figure 1).  {\em The
properties of the point sources and of the diffuse emission are strongly
correlated.} 
\item{} Consistent with earlier results, none of the point sources
detected here is positionally coincident with the strong, compact OH maser
features in Arp 220.  (However, several sources show both OH absorption
and low-gain maser amplification, results in preparation).
\end{itemize}

We can also compare the current results to earlier data.  The most
relevant comparison at this time is to data taken in experiment GL26B, 12
months earlier than the current GD17A observations.  The GL26B data have
presented challenging calibration problems, but an image of sufficient
quality for meaningful comparison, with rms noise of $\sim9\mu$Jy, is
available.  We find that in the intervening 12 months, at least 4 new
sources have appeared, namely E8, W11, W25 and W34.  Their flux densities
in the current image are such that they would have been readily detected
in the GL26B data.  Source W40 is also possibly new, but the evidence is
less compelling.  Another, bright source (W12), was present 12 months
earlier, but conspicuously absent 3.4 years earlier (by comparison with
data from Rovilos et al., 2005).

It is premature to derive new, high sensitivity light curve data for a
large number of sources.  The necessary monitoring data have been taken
for multiple epochs, but the analysis is complex and time-consuming, and
accurate flux scale registration requires additional work.  These results,
including the GL26B data referred to above, will be presented separately.

\subsection{Radio Supernovae and the Nuclear Starburst}

The original interpretation of the point sources in Arp 220 as radio
supernovae (RSNe) by Smith et al. (1998a) remains the favored one.  The
sources are unresolved, and exhibit flux density variations on timescales
of months and years (Rovilos et al. 2005).  Several new sources have been
observed to appear where none previously existed.  The observed flux
densities, between 50 and 1200 $\mu$Jy, correspond to 1.6 GHz radio powers
between 3.5$\times 10^{19}$ and 8.3$\times 10^{20}$W Hz$^{-1}$, which
falls in the upper end of the range of observed peak powers for known RSNe
(van Dyk et al., 2000).

The widespread variability, and frequent appearance of new sources
constitutes compelling evidence that a significant fraction of the sources
are young and compact.  Is there, however, a possibility that some of the
sources are much older, and would be more conventionally classified as
supernova remnants (SNR)?  There are, however, no known examples of
decades or centuries-old SNRs with radio luminosities approaching those of
the sources in Arp 220.  The faintest detected sources in Arp 220 are of
order 50 times more luminous than the brightest galactic SNRs with known
explosion dates (e.g. Cassiopeia A), and typically an order of magnitude
more luminous than the bright resolved RSN in the nearby starburst galaxy
M82 (see next subsection).  The most plausible interpretation is that all
the detectable sources in Arp 220 are young, and consequently luminous.
Given the apparent rate of appearance of new sources, and the total number
of sources detected, the turnover rate of sources in a uniform population
would imply that the oldest objects are likely to be at most a few
decades old.  However, the population is clearly heterogeneous, and there
is evidence that the brighter sources tend to decay slowly.  More
information is needed in order to reliably estimate ages.

Recently, our team has detected several of the point sources at 2.3, 5
and 8.4 GHz (Conway et al, in prep); preliminary analysis indicates
that they exhibit a range of spectral indices, from flat to
steep. Further analysis and interpretation will be provided in the
forthcoming paper. In the absence of external free-free absorption,
RSN which are no longer in the rising phase at 18cm are expected to
exhibit steep spectra characteristic of optically thin synchrotron
emission.  There is considerable evidence, however, for significant
free-free optical depths at 18cm in the nuclei of powerful ULIRGs
(e.g. Condon et al. 1991), which would likely manifest itself as
spatially correlated flattening of RSN spectra across regions of the
Arp 220 nuclei.  Any interpretation of the star-forming environment in
the nuclei, based on our 18cm-only RSN measurements, must take into
account the possibility that radio luminosities, RSN counts, and rates
of new source appearances may all be underestimated.

\subsubsection{Differing RSN populations in east and west}

The two nuclei of Arp 220, though strikingly similar in many properties
(numerous RSN, strong compact OH megamaser emission, strong diffuse
continuum emission), display differing properties with regard to the point
sources.  The two principal differences are: (a) the point sources are
systematically stronger in the west (i.e. the luminosity functions are
different as can be seen in Figure 3); and, (b) the sources are confined to
a somewhat more compact region in the west.  We can quantify the
difference in luminosity functions by applying a Kolmogorov-Smirnov test
to the flux density data.  A simple comparison yields a probability of
0.001 that the samples are drawn from the same population, primarily due
to the much higher mean flux density in the west. 

An obvious hypothesis to explain this result is that the eastern nucleus
population is just a fainter replica of that in the west.  A simple model
relating expected RSN luminosities to the far infrared luminosity can be
constructed, using a $t^{-\gamma}$ decay profile where $\gamma = 1.3$
(Smith et al. 1998a,b), and translating $L_{fir}$ into a relative star
formation rate and hence a supernova rate.  This leads to mean RSN ages
proportional to $L_{fir}^{-1}$, and RSN fluxes $S_{rsn} \propto
L_{fir}^{\gamma}$.  The fainter sources in the east would then simply
reflect a less-luminous starburst than in the west, which is qualitatively
consistent with the less compact morphology.  Note that such a model
cannot modify the shape of the RSN luminosity function, but can only
translate it to lower luminosities as the starburst intensity and
associated $L_{fir}$ declines.  

To test this, we can truncate the western source list to the brightest 20
objects, and scale the fluxes to match the effective sensitivity limit in
the eastern nucleus - reduction by a factor of 2.7.  Comparing the
resulting distributions, the K-S test indicates a probability of
$\sim$13\% that the samples are drawn from the same population - the
populations still differ, though the confidence level is not high.  It
appears that a simple model in which the western nucleus hosts a higher
star formation rate than the eastern nucleus, for whatever reason, may not
adequately predict the observed point source properties.  This suggests
that other factors may be involved, such as differing chemical
compositions and IMFs between the nuclei resulting in a different mix of
supernova types, differing ISM densities influencing the RSN light curves,
or a time dependence in the exponent $\gamma$ of unknown origin.

\subsubsection{Comparison with SN 2000ft in NGC7469}

Recently, a luminous RSN has been reported in the ULIRG NGC7469 (Colina et
al., 2001, Alberdi et al. 2006).  This source, SN 2000ft, exhibits a
well-determined classic type II RSN light curve, with an accurately
modeled explosion date and a relatively steep flux density decay ($\gamma
= 1.8$).  NGC7469 is at a similar distance to Arp 220, and SN 2000ft
reaches modeled peak 8GHz, 5GHz and 1.6 GHz flux densities of 2 mJy, 1.5
mJy and 0.35 mJy respectively.  In Arp 220, this supernova would thus have
appeared at 1.6 GHz as a moderately bright but very short-lived RSN among
a population that is typically much longer-lived.  Alberdi et al. argue
that the classical behaviour of the RSN indicates a light curve governed
by ejecta interacting with the pre-existing stellar wind (e.g. Weiler et
al. 1990).  They predict that when the ejecta reach the dense ISM presumed
to surround the supernova site, the light curve decay rate will slow
dramatically.  

Clearly, many of the RSN in Arp 220 are behaving differently, and the slow
decay (ISM interaction) phase appears to have been reached while the radio
luminosity is still high.  This would be a natural consequence if these
objects are embedded in a denser ISM than SN 2000ft, so that the ISM
interaction phase occurs earlier.  Such may commonly be the case for RSN
in the compact nuclei of Arp 220, which are likely to have a higher mean
density than the 1 kpc-scale circumnuclear starburst region of NGC7469.  

This comparison leads to the expectation of a range of behaviours
within the Arp 220 RSN population, reflecting ISM inhomogeneity, and a
range in the strength and onset timing of the ISM interaction phase.
Heterogeneity in the RSN population renders efforts to constrain the
supernova rate by examination of luminosity functions and light curves
unattractive.  The supernova rate is instead best determined directly via
sufficiently sensitive and frequent monitoring for new sources.  It can
then be used as a powerful constraint on models in order to understand the
evolution of the population.

\subsubsection{Supernova rate and starburst luminosity}

By virtue of a severalfold improvement in sensitivity and RSN detection
threshold, we have directly observed 4 new sources in a 12-month period,
with three in the western nucleus and one in the eastern nucleus.  We do
not know whether all type-II supernovae in Arp 220 maintain detectable
flux density levels within 12 months of initial appearance, but if so, we
have a measure of the supernova rate in the galaxy at 4$\pm$2 per year.
If not, we may have missed some, and this becomes a lower limit.

Smith et al. (1998a,b) discuss the relationship between FIR luminosity,
star formation rate, the stellar IMF, and the supernova rate.  They
conclude that based on the luminosity of Arp 220, a supernova rate of
~2/yr can be expected, with an uncertainty of about a factor of 2
depending primarily on assumptions about the IMF.  Our estimate of 4$\pm
2$/yr based on the current observations is consistent with the Smith et
al.  estimate.  Clearly, the starburst traced by these RSNe is capable of
generating the entire FIR luminosity of Arp 220, within the uncertainties
of our measurements and of starburst modelling.  A hidden AGN, while not
ruled out by these observations, is not required on energetic grounds.

A high star formation and supernova rate in Arp 220 would be expected to
generate a large X-ray luminosity from X-ray binaries.  A relationship
between starburst-powered FIR luminosity and X-ray luminosity has been
established (e.g. Persic et al. 2004), but Arp 220 does not fall on the
correlation, being underluminous in the X-rays.  This could be explained
by extreme absorption column densities, or by relative youth of the
starburst compared to XRB lifetimes resulting in a deficiency of XRBs.
Imanishi et al. (2005) report XMM-Newton detection of prominent Fe
K$\alpha$ line emission, throwing into doubt the interpretation of the
weak hard X-ray emission in Arp 220 as originating from X-ray binaries,
and further exacerbating the problem.  They appealed to the possibility of
a low supernova rate and an energetically dominant buried AGN to explain
their findings.  The current work instead strongly confirms a high
supernova rate, so the X-ray binary issue remains problematic.  A robust
independent estimate of typical absorption columns to the starbursting
regions of Arp 220 is needed, and the measurement of free-free optical
depths through VLBI techniques will be valuable in this regard.

\subsubsection{West-east luminosity ratio} 

An important question concerns the ratio of luminosities between the two
nuclei.  Using 25$\mu$m imaging, Soifer et al. (1999) measured a 3:1
ratio, with the western nucleus being the stronger.  Since most of the
luminosity of Arp 220 emerges at longer wavelengths (around 100$\mu$m),
and since there is evidence for significant opacity at 25$\mu$m (Soifer et
al. 2002), the infrared evidence strongly suggests a 3:1 luminosity ratio,
but is not conclusive, and does not dictate an origin (AGN or starburst)
for the luminosity.

The RSN data presented here support a roughly 3:1 ratio of starburst
luminosities between the nuclei.  The different rates in the east and the
west (1/yr and 3/yr respectively), while obviously very poorly determined
due to small number statistics, are roughly consistent with ratio of 
luminosities (2.7) assumed in 4.2.1 for testing the hypothesis that the
eastern nucleus differs from the west only in starburst intensity and
typical RSN ages.  Future monitoring will quickly refine this statistical
result.

The apparent agreement between infrared and RSN evidence for a 3:1 ratio
lends weight to the hypothesis that the bolometric luminosity of Arp 220
is dominated by starburst activity.  If AGN activity were a major
contributor, one would expect the overall ratio to differ from the ratio
based solely on starburst activity.  Although requiring confirmation with
more data unaffected by opacity, this appears not to be the case;
starburst activity traced by radio supernovae yields the same luminosity
ratio as a method (IR imaging) which is blind to the luminosity source.

\subsubsection{The Diffuse Radio Continuum}

A supernova origin for the synchrotron plasma responsible for the diffuse
continuum emission is strongly supported by our data.  We see a detailed
correspondence between the properties of the supernova aggregations and
the diffuse emission regions, consistent with diffusion of supernova
remnant relativistic electrons into the interstellar medium.  The degree
of correspondence suggests that the electrons are not diffusing very far,
most likely due to relatively short synchrotron lifetimes of the
electrons.  The ISM density could also be involved, since it may influence
both the local star formation rate and the local synchrotron emissivity
when seeded with SNR electrons.

This emission is of the kind expected to conform to the well-known
FIR/radio correlation, and indeed it has been shown that ULIRGs in general
follow this correlation, after correction for free-free absorption (Condon
et al.  1991).  We therefore expect an underlying bolometric luminosity
ratio of 3:1 between the nuclei to be reflected in a similar ratio of
diffuse continuum fluxes.  This does not appear to be the case; at 18cm
the diffuse continuum flux ratio between the two nuclei is only $\sim$1.2
(Rovilos et al.  2003).  Further, Rodriguez-Rico et al. (2005) measure a
ratio of $\sim$1.25 at both 8.4 GHz and 43 GHz.  A ratio of closer to 2:1
at the emission peaks is indicated from observations at higher angular
resolution by Rovilos et al. (2003) at 5 GHz, and by Norris (1988) at 5,
15 and 22 GHz.

This apparent discrepancy can be understood through a combination of
differential free-free absorption at 1.6 GHz, and differential energy loss
rates in the synchrotron plasma.  The western nucleus is stronger and more
compact than the eastern nucleus.  In fact, in our 1.6 GHz VLBI
observations, the diffuse emission from the western nucleus is
sufficiently compact at its peak that it is strongly detected on our
shortest baselines, which must be excised in order to preserve imaging
fidelity of the RSN (see section 2).  The diffuse continuum brightness
distribution is strongly peaked, and is poorly modelled by a Gaussian.
This compactness of the peak will be associated with higher free-free
optical depths, and higher energy densities and pressures which will
plausibly lead to more rapid expansion losses in the synchrotron plasma.

Regarding free-free absorption, following Smith et al. (1998b) the mean
absorption optical depth is likely to be $\sim$0.5 in Arp 220, and
certainly higher in the inner regions of the western nucleus.  This is
adequate to explain the 1.6 GHz result of Rovilos et al. (2003).

To explain the low radio flux ratio at higher frequencies where free-free
absorption is negligible, we must consider the properties and evolution of
the diffuse synchrotron plasma.  It is noteworthy that the observed
diffuse radio properties of the Arp 220 nuclei yield lifetimes for
electrons radiating at $\sim$5 GHz of between $10^4$ and $10^5$ yr
(assuming field/particle energy equipartition).  These times are short
compared to typical starburst lifetimes of 10$^7$ yr or more, so in the
absence of other energy loss mechanisms, a steady-state system will
result, with synchrotron luminosities mirroring the electron injection
rates from supernovae.  However, the evidence is compelling that in
powerful nuclear starbursts like that in Arp 220, supernovae generate
local overpressures which drive superwinds.  These superwinds are often
detectable in X-rays, and evidence for a powerful superwind has been found
in Arp 220 itself (Heckman et al. 1996, McDowell et al. 2003).  Since the
western nucleus exhibits direct evidence for a higher mechanical energy
injection rate per unit volume from supernovae than the eastern nucleus
(this paper), it is plausible that the pressure is correspondingly higher,
and the resulting expansion losses more rapid, compared to the eastern
nucleus.  Such differential expansion losses would depress the diffuse
radio continuum flux density in the west, as observed.

Free-free absorption, synchrotron aging and expansion losses will all
manifest themselves in the spectral behavior of the radio emission from
the two nuclei.  Detailed high-sensitivity spectral mapping at 0.1
arcsecond resolution will be needed to further constrain these phenomena
in the Arp 220 nuclei.  This may soon be possible with the expanded
bandwidths and sensitivity of the e-MERLIN instrument.

On a cautionary note, free-free absorption will preferentially depress the
flux densities of the RSN in the western nucleus.  If these reduced flux
densities fall below our detection limit for some new sources up to a year
old, the inferred supernova rate in the western nucleus will be depressed
relative to that in the east.  The only way to discover if our census of
new supernovae is complete, and therefore if our inferred supernova rate
estimates are unbiased between the two nuclei, is to see if increased
imaging sensitivity yields increased rates of appearance of new sources.
This will be feasible in coming years as bandwidths for VLBI systems
continue to improve.

To summarize our conclusions from this section, our images of RSNe trace
an energetic starburst in progress in both Arp 220 nuclei.  Regarding the
luminosity ratio between the nuclei, comparison of our data with mid-IR
imaging data leads to a consistent picture of a 3:1 ratio of west/east
luminosities.  A starburst origin for the bulk of the Arp 220 bolometric
luminosity is supported by this comparison.  The failure of the diffuse
radio continuum emission to exhibit a similar 3:1 ratio can be understood
in terms of differential free-free absorption, and more speculatively, in
terms of differential synchrotron plasma expansion losses.

\subsection{Comparison to M82}

The starburst galaxy M82 lies at a distance of only 3.2 Mpc, and as such
is amenable to detailed study at much lower luminosity levels.  It is the
only other galaxy in which large numbers of compact sources have been
detected.  Despite the obvious difference in the scale of the phenomenon,
there are striking similarities, and we deem a comparison of the two
galaxies to be useful.  The far infrared luminosity of M82 is 2.8$\times
10^{10}L_{\odot}$ (Telesco 1988), which is roughly 50 times lower than
that of Arp 220.  The star formation rate is therefore expected to be
lower by a comparable factor, and with a consequently lower supernova
rate, one expects to see a population of supernova remnants that is much
older and fainter than that in Arp 220.  Qualitatively, this is exactly
what is found.

In Figure 3 we show a histogram of compact radio source monochromatic 18cm
power for M82, and for each of the two nuclei of Arp 220.  The detection
threshold for Arp 220 is  $\sim 10^{19.6}$W Hz$^{-1}$, while that for
M82 is $\sim 10^{18.0}$W Hz$^{-1}$.  It is notable that only one of the
M82 sources falls above the Arp 220 luminosity detection threshold.  If
the supernova rate in M82 is 50 times lower than in Arp 220, consistent
with the ratio of FIR luminosities, we expect only one RSN in M82 of
comparable youth to the ~50 that we have detected in Arp 220.  In fact,
the last RSN to appear in M82 was $\sim$40 years ago, so another is
somewhat overdue.  The bright M82 source 41.95+57.5 is decaying at
$\sim$8.5\%/yr, and while it may be a plausible candidate for a young RSN,
there is strong evidence that this object is atypical, may not be an RSN,
and may be much older (Pedlar et al. 1999, McDonald et al. 2001, Beswick
et al. 2006 submitted).  The hypothesis that the starburst in M82 is a
scaled-down version of the same phenomenon occurring in Arp 220 is thus
attractive, and supported by the data.

Based on this, we can make an estimate of the number and strength of
supernova remnants, similar to those observed in M82, that lie below our
detection threshold in Arp 220.  This number should scale with the FIR
luminosity, and the resulting total flux density from ~1500 SNRs is
$\sim$11 mJy, comparable to the total flux density in the 49 detected
sources.  Based on the observed sizes of the M82 sources (e.g. Muxlow et
al. 1994, McDonald et al. 2001), the SNRs will have a volume filling
factor of up to a few percent in the Arp 220 nuclei.  The predicted 11 mJy
will contribute to the apparently diffuse emission from the Arp 220
nuclei, but the flux density will actually reside in angularly compact but
undetectably weak supernova remnants.  This is in addition to the $\sim$12
mJy of the nuclear flux density that resides in the detected sources.
Together, these total $\sim$12\% of the nuclear and $\sim$8\% of the total
18cm flux density in Arp 220.  This constitutes a measure of the number of
relativistic electrons that remain trapped in high-emissivity regions of
supernova remnants, and that have not yet diffused into the lower
emissivity environment of the general ISM.

This conclusion must be tempered by the realization that the evolution
of older supernova remnants in the dense Arp 220 environment may differ
systematically from that occurring in M82.

While the hypothesized weak SNRs may be undetectable in continuum
emission, it is possible that they will contribute compact spots of
enhanced continuum brightness, which may exhibit OH maser amplification to
detectable levels.  Interpretation of OH maser properties of galaxies such
as Arp 220 should take this possibility into account.

\section{Conclusions}

We have presented a $\lambda$18cm VLBI image of the compact sources in Arp
220, with $\sim$1pc resolution and unprecedented sensitivity.  A robust
analysis of the image has yielded detections of 49 unresolved sources, of
which 20 lie in the eastern nucleus, and 29 lie in the west.  The eastern
sources are systematically weaker than those in the west.  There is some
evidence that in addition, the shape of the luminosity distributions
differs between the two nuclei.  Four new sources have been observed to
appear over a 12-month period, with one in the east and three in the west.

The interpretation of these sources as radio supernovae (RSNe) appears secure,
and is strengthened by the appearance of new sources on a short timescale.
The detected RSNe population in Arp 220 spans the upper end of the range
of peak 18cm radio powers for type II supernovae.  The rate at which new
sources are appearing yields an estimate of the supernova rate of 4$\pm 2$/yr.
There remains a possibility that this rate is underestimated due to
sensitivity limitations.  The supernova rate can be related to the star
formation rate, and current data imply that star formation activity in Arp
220 is sufficient to power the entire observed far infrared luminosity of
the galaxy.

The differences between the two nuclei of Arp 220 are not yet fully
understood.  However it is clear that the western nucleus is more compact,
and evidence from both IR imaging and the current RSN results indicates
that it is $\sim$3 times more luminous than the eastern nucleus.
Differential free-free absorption and expansion losses in the synchrotron
plasma may be responsible for the failure of the diffuse continuum
emission to exhibit a similar 3:1 flux ratio between the nuclei.  There
appear to be additional differences between the RSN populations in the two
nuclei that have yet to be explained.  It is expected that the ISM will
strongly influence the observed RSN properties, and the ISM may differ
between the nuclei.  Measurements of free-free absorption to each RSN
source via multi-frequency VLBI will help to distinguish among
possibilities.

A comparison of Arp 220 with the nearby, well-studied starburst galaxy M82
yields a consistent picture with regard to supernova-related compact radio
sources, in which the Arp 220 starburst is simply $\sim$50 times more
luminous, and is confined to a smaller volume.  By analogy with M82,
roughly 6\% of the apparently diffuse continuum emission from the Arp 220
nuclei may reside in an ensemble of weak, compact supernova remnants.
This may have observable consequences for OH maser emission
characteristics.

\acknowledgments
The National Radio Astronomy Observatory is a facility of the National
Science Foundation operated under cooperative agreement by Associated
Universities, Inc.  The Arecibo Observatory is part of the National
Astronomy and Ionosphere Center, which is operated by Cornell University
under a cooperative agreement with the National Science Foundation.  The
European VLBI Network is a joint facility of European, Chinese, South
African and other radio astronomy institutes funded by their national
research councils.  This work was partially supported by NSF grant
AST-0352953 to Haystack Observatory.  We thank Rob Beswick for help in
compiling data on M82, and useful comments on the manuscript.

\clearpage

\begin{landscape}
\begin{figure}
\includegraphics[scale=0.2]{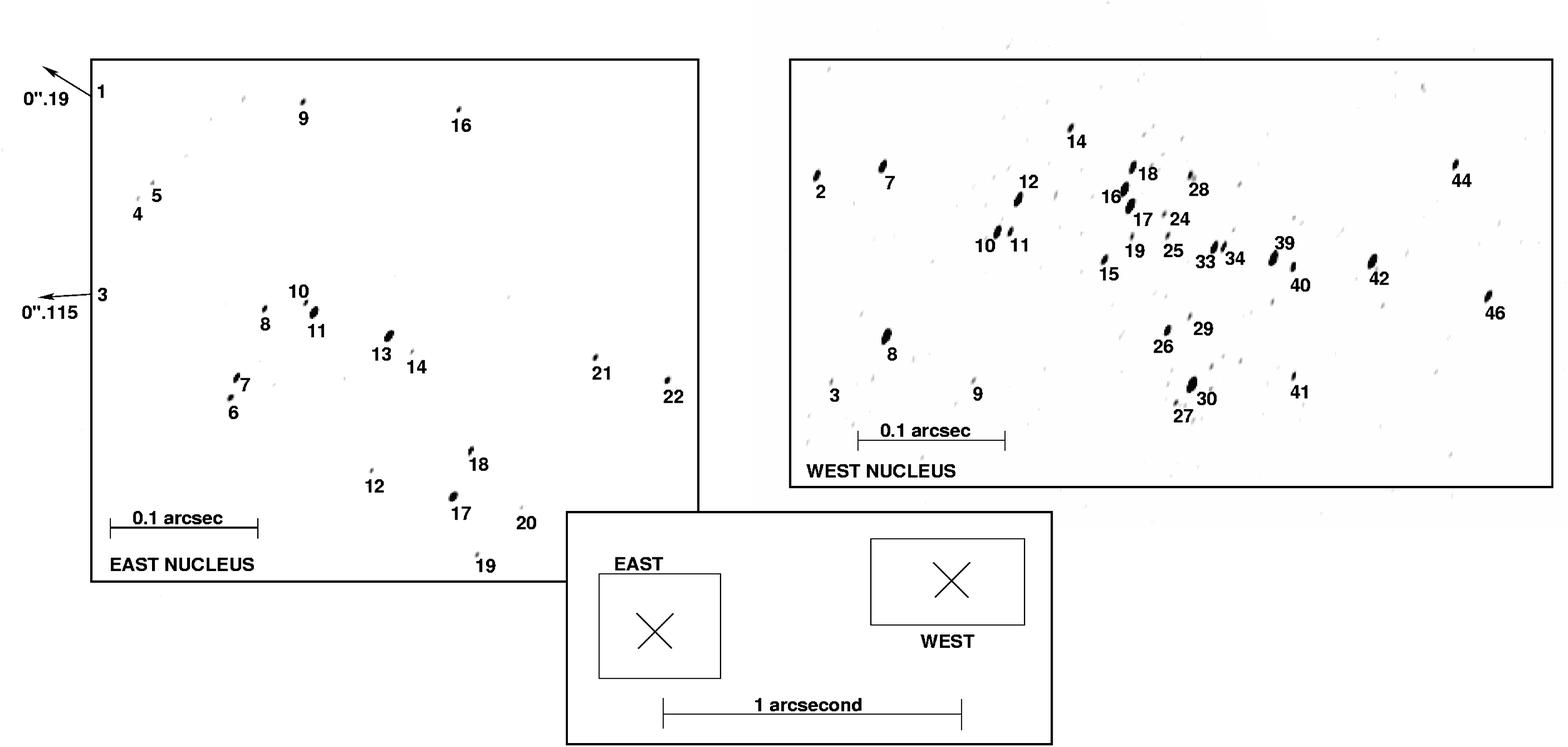}
\caption{\small Continuum images of the east and west nuclei of Arp 220.  The
angular resolution is 5.9$\times$2.7 milliarcsecond in position angle
-21$^{\circ}$.  The background rms level in the east nucleus ranges from 5
to 6 $\mu$Jy/beam, and in the west nucleus ranges from 7 to 9
$\mu$Jy/beam.  Sources which meet the detection criteria (see text) are
numbered separately for east and west nuclei, and are identified in Table
1 as E{\em n} and W{\em n} respectively, where {\em n} is the source
number.  The inset shows the relative locations of the two images, and the
crosses mark the relative locations of the peaks of the diffuse radio
continuum emission in the two nuclei, as measured using MERLIN at 6cm
(Figure 2 of Rovilos et al., 2003).  It can be seen that the point source
locations are highly correlated with the diffuse continuum distribution.
The diffuse emission, which is ~20 times stronger than the sum of the
point source fluxes, is fully resolved out by our VLBI array.}
\end{figure}
\end{landscape}

\begin{figure}
\epsscale{.80}
\plotone{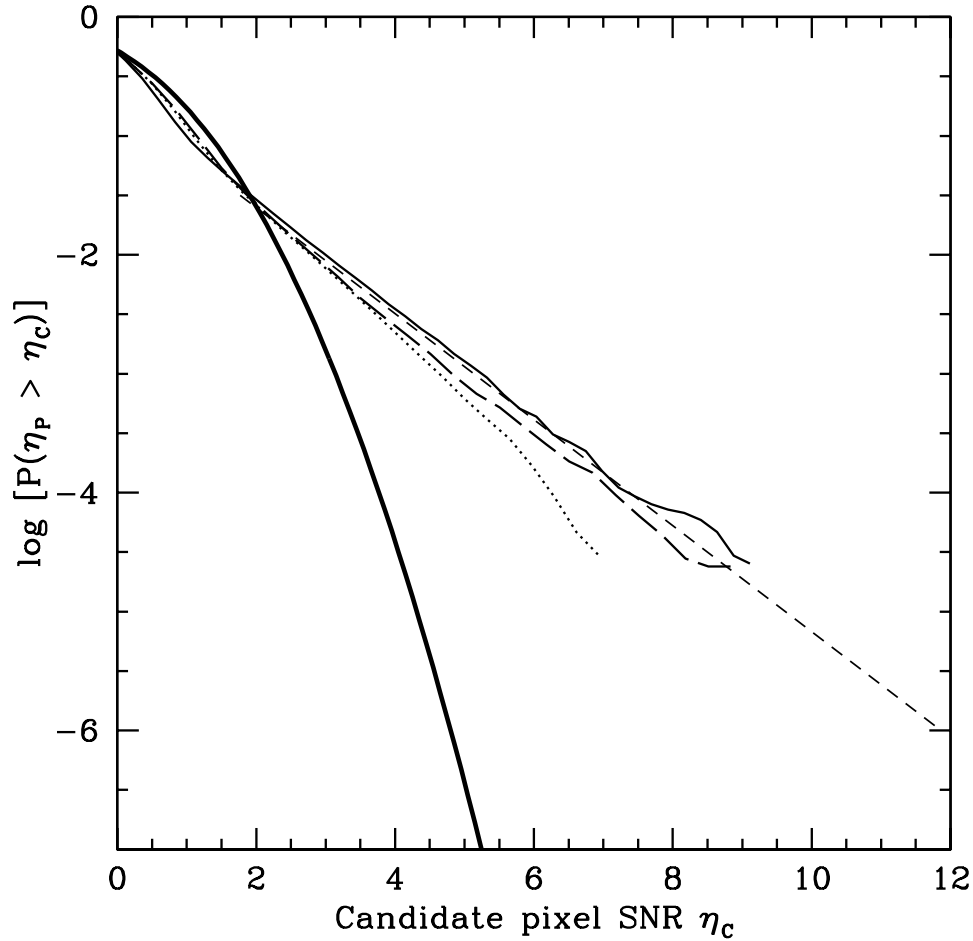}
\caption{Noise statistics for various regions of the Arp 220 image.  The
horizontal axis is SNR of a candidate pixel, $\eta_C$, defined as the
measured value divided by the local rms.  The vertical axis gives the
probability that a randomly selected pixel of SNR $\eta_P$ will exceed a
given candidate pixel SNR $\eta_C$, based on the statistics of the image
in a region.  The long-dash curve refers to the eastern nucleus, the
solid curve to the western nucleus, and the dotted curve to a region
between the two nuclei.  The short-dash straight line is a linear least
squares fit to the curves for the east and west nuclei, and is the basis
for extrapolation to observed SNR values of source candidates, and thence
to estimates of the probability of false detection on a case-by-case basis
(see text).  The heavy black curve represents a Gaussian distribution of
pixel values.} 
\end{figure}

\begin{figure}
\plotone{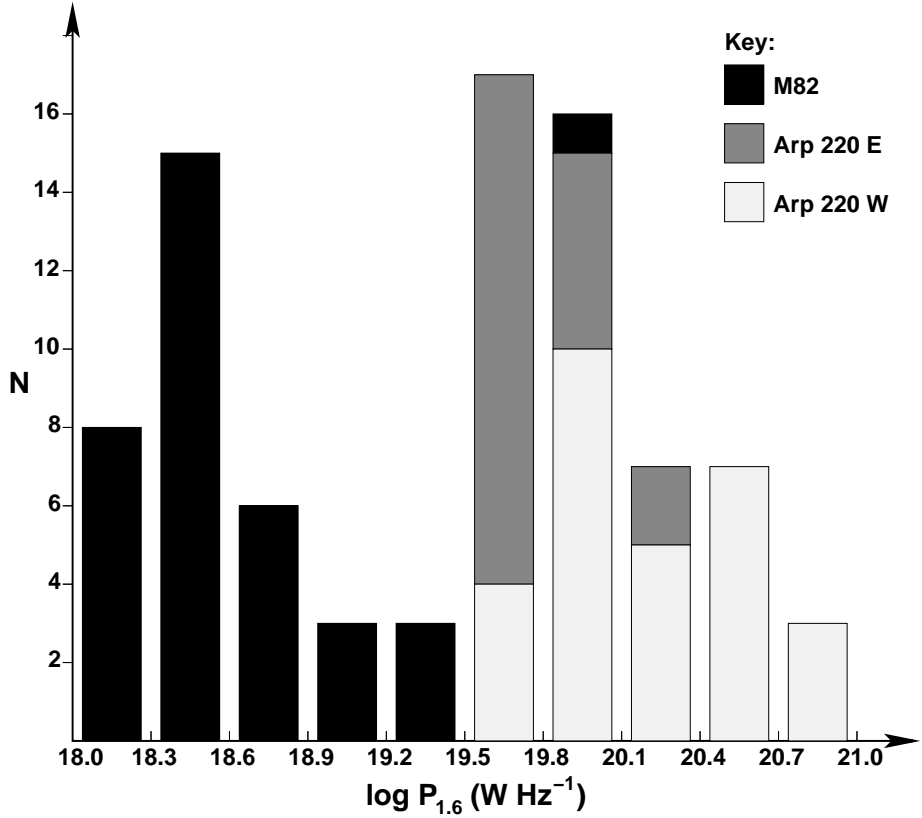}
\caption{Histogram of the 1.6 GHz monochromatic power for compact sources
in M82 and Arp 220.  There are two important features illustrated here.
First, the M82 sources are almost all weaker than those detected
in Arp 220, with typical Arp 220 sources being several tens of times
stronger than typical M82 sources.  The exception to this is the source
41.95+57.5 in M82, which is declining at 8.5\%/yr (Kronberg et al., 2000).
Second, the power distributions of sources in the E and W nuclei of Arp
220 are markedly different.  Not only are the eastern sources weaker on
average, but the shape of the distribution appears more strongly biased
toward weaker sources.  The M82 data are taken from Wills et al. (1998),
Allen and Kronberg (1998), and private communication from R.  Beswick.
The Arp 220 data are from this paper.} 
\end{figure}

\clearpage

\clearpage

\begin{deluxetable}{lllll||lllll}
\tabletypesize{\scriptsize}
\tablecaption{Point source candidates in Arp 220}
\tablewidth{0pt}
\tablehead{
\colhead{Name} & \colhead{$\delta$ RA (mas)} & \colhead{$\delta$ Dec. (mas)} 
    & \colhead{Flux ($\mu$Jy)} & \colhead{$\Psi_C$} & 
\colhead{Name} & \colhead{$\delta$ RA (mas)} & \colhead{$\delta$ Dec. (mas)} 
    & \colhead{Flux ($\mu$Jy)} & \colhead{$\Psi_C$} } 
\startdata
E1  & 658.3  & 225.3  & 69.7   & 0.04  &  W9  & -422.3 & 50.0   & 74.5   & 0.00 \\
E3  & 617.4  & -18.8  & 70.6   & 0.01  &  W10 & -438.5 & 150.6  & 408.8  & 0.00 \\
E4  & 496.9  & 56.1   & 52.9   & 0.00  &  W11 & -447.2 & 150.9  & 175.4  & 0.00 \\
E5  & 488.3  & 67.3   & 57.3   & 0.14  &  W12 & -452.8 & 173.0  & 645.6  & 0.00 \\
E6  & 442.3  & -87.3  & 90.8   & 0.00  &  W14 & -487.6 & 221.2  & 135.9  & 0.00 \\
E7  & 438.6  & -72.8  & 135.2  & 0.00  &  W15 & -510.5 & 131.8  & 189.4  & 0.00 \\
E8  & 422.1  & -23.4  & 92.6   & 0.00  &  W16 & -524.0 & 179.9  & 655.4  & 0.00 \\
E9  & 399.5  & 125.6  & 81.1   & 0.00  &  W17 & -527.7 & 168.4  & 717.4  & 0.00 \\
E10 & 397.8  & -19.0  & 80.0   & 0.01  &  W18 & -529.2 & 194.4  & 255.1  & 0.00 \\
E11 & 392.9  & -26.0  & 296.3  & 0.00  &  W19 & -528.8 & 147.5  & 95.5   & 0.10 \\
E12 & 358.8  & -139.8 & 65.4   & 0.03  &  W24 & -550.6 & 163.0  & 90.0   & 0.06 \\
E13 & 348.7  & -43.0  & 234.8  & 0.00  &  W25 & -552.9 & 148.2  & 98.0   & 0.01 \\
E14 & 334.8  & -54.2  & 62.1   & 0.09  &  W26 & -552.6 & 83.9   & 219.8  & 0.00 \\
E16 & 307.2  & 120.1  & 79.6   & 0.00  &  W27 & -558.2 & 35.1   & 96.4   & 0.03 \\
E17 & 310.5  & -158.7 & 139.3  & 0.00  &  W28 & -568.0 & 189.3  & 116.2  & 0.00 \\
E18 & 300.1  & -125.4 & 99.4   & 0.00  &  W29 & -567.6 & 93.3   & 85.6   & 0.07 \\
E19 & 296.6  & -200.2 & 64.2   & 0.05  &  W30 & -569.0 & 47.0   & 1228.0 & 0.00 \\
E20 & 270.2  & -166.4 & 52.0   & 0.14  &  W33 & -584.1 & 140.3  & 428.0  & 0.00 \\
E21 & 226.4  & -58.3  & 86.0   & 0.00  &  W34 & -590.6 & 140.7  & 160.8  & 0.00 \\
E22 & 183.9  & -74.8  & 97.4   & 0.00  &  W39 & -624.0 & 133.0  & 870.2  & 0.00 \\
    &        &        &        &       &  W40 & -637.0 & 127.1  & 195.5  & 0.00 \\
W2  & -317.0 & 189.0  & 253.6  & 0.00  &  W41 & -637.5 & 53.2   & 111.7  & 0.00 \\
W3  & -327.0 & 49.4   & 75.3   & 0.04  &  W42 & -690.5 & 131.0  & 713.5  & 0.00 \\
W7  & -361.6 & 195.3  & 368.8  & 0.00  &  W44 & -746.2 & 196.6  & 207.5  & 0.00 \\
W8  & -363.8 & 79.8   & 849.3  & 0.00  &  W46 & -768.0 & 107.2  & 271.2  & 0.00 \\
\enddata

%% Text for table notes should follow after the \enddata but before
%% the \end{deluxetable}. Make sure there is at least one \tablenotemark
%% in the table for each \tablenotetext.

\tablecomments{Table 1.  Source candidates and measured properties in rough
right ascension order.  A source candidate is included in the table if its
flux density exceeds 9.5 times the local rms of the background (with the
exceptions of E4 and W9 as discussed in the text).  Column 5
lists the probability of false detection ($\Psi_C$) of each candidate based on
analysis of the noise statistics of the image, as defined in the text.
The sum of these probabilities is such that the expectation for the number
of false positive detections is less than one, out of this list of 49
candidates.}
\end{deluxetable}

\end{document}